\documentclass[11pt,amsfonts]{amsart}

\input fullpage.sty

\newtheorem{theorem}{Theorem}[section]

\newtheorem{corollary}[theorem]{Corollary}

\newtheorem{definition}[theorem]{Definition}

\newtheorem{remark}[theorem]{Remark}

\def\endproof{\qed \medskip}
\def\blacksquare{\hbox to .60em{\vrule width .60em height .60em}}

\begin{document}

\title[Regularity for Lorentz Metrics]{Regularity for Lorentz Metrics 
under Curvature Bounds}

\author[M. Anderson]{Michael T. Anderson}

\thanks{Partially supported by NSF Grant DMS 0072591}

\maketitle

\abstract 
Let ({\bf M, g}) be an $(n+1)$-dimensional space time, with bounded 
curvature, with respect to a bounded framing. If ({\bf M, g}) is 
vacuum, or satisfies a mild condition on the stress-energy tensor, then 
we show that ({\bf M, g}) locally admits coordinate systems in which 
the Lorentz metric {\bf g} is well-controlled in the (space-time) 
Sobolev space $L^{2,p}$, for any $p < \infty$.
\endabstract

\setcounter{section}{0}
\setcounter{equation}{0}

\section{Introduction}

 A well-known issue in the geometry of space-times is to understand the 
regularity of metrics with given bounds on the curvature tensor. This 
issue arises frequently in discussions and analysis of the behavior at 
the boundary and definitions of singularities for space-times, c.f. 
[11], [6], [7], [14] for example. 

 More specifically, it has been an open problem for some time, cf. 
[4]-[6], [14] for instance, whether a space-time ({\bf M, g}) which has 
curvature bounded in $L^{\infty}$ in a suitable sense has coordinate 
charts in which the metric 
${\bf g} = {\bf g}_{\alpha\beta}$ is $C^{1,\gamma}\cap L^{2,p},$ for 
any $\gamma  < 1$, $p <  \infty$. Here $C^{k,\gamma}$ is the H\"older 
space of functions whose $k^{\rm th}$ derivatives are H\"older 
continuous of order $\gamma$, while $L^{k,p}$ is the Sobolev space of 
functions with $k$ weak derivatives in $L^{p}.$

 The purpose of this paper is to provide an affirmative solution to 
this problem, at least for vacuum space-times or space-times satisfying 
a mild condition on the stress-energy tensor.

\medskip

 The solution of the corresponding problem in Riemannian geometry has 
been known for some time, and it is useful to state the exact result in 
this context before considering the Lorentzian analogue. Thus, let $(M, g)$ 
be a Riemannian $n$-manifold, with say $C^{\infty}$ smooth metric 
$g$. Suppose there exists a point $p\in M$ such that
\begin{equation} \label{e1.1}
dist_{g}(p, \partial M) \geq  1. 
\end{equation}
Let $R = R_{ijkl}$ denote the Riemann curvature tensor of $(M, g)$, and 
let $B_{p}(r)$ denote the geodesic ball of radius $r$ about $p$ in 
$(M, g)$. Suppose one has bounds
\begin{equation} \label{e1.2}
|R|_{L^{\infty}(B_{p}(1))} \leq  C,  \ \ vol_{g}B_{p}(\tfrac{1}{2}) 
\geq  v_{o}, 
\end{equation}
for arbitrary constants $C <  \infty , v_{o} > $ 0. Then there exists a 
constant $r_{o} > $ 0, depending only on $C$, $v_{o}$ and $n$, such 
that the ball $B_{p}(r_{o})$ admits a coordinate chart $U = \{u_{k}\},$ 
in which the metric $g$ is $C^{1,\gamma}\cap L^{2,p},$ for any 
$\gamma < 1$, $p <  \infty$. Further, there exists a constant $R_{o},$ 
depending only on $C$, $v_{o}$, $n$ and $p$, such that
\begin{equation} \label{e1.3}
||g_{ij}||_{L^{2,p}} \leq  R_{o}, 
\end{equation}
where the norm is taken over the ball $B_{p}(r_{o}).$ A proof of this 
result may be found in [12] for instance. By Sobolev embedding 
$C^{1,\gamma} \subset  L^{2,p},$ for $\gamma  = 1-\frac{n}{p}$, so that 
(1.3) also gives a bound on $g_{ij}$ in $C^{1,\gamma}.$

\medskip

 A direct analogue of this result in Lorentzian geometry is false, due 
to the existence of large families of non-flat space-times for which 
the curvature norm $|{\bf R}|^{2} = {\bf R}_{ijkl}{\bf R}^{ijkl}$ 
vanishes identically. Thus, consider for instance the class of vacuum 
plane-fronted gravitational waves on ${\mathbb R}^{4},$ with metric of 
the form
\begin{equation} \label{e1.4}
{\bf g} = -dudv - h(x,y,u)du^{2} + (dx^{2} + dy^{2}), 
\end{equation}
\begin{equation} \label{e1.5}
\Delta_{(x,y)}h = 0. 
\end{equation}
For such metrics, the two possible scalar invariants in the curvature 
tensor, namely
$$|{\bf R}|^{2} = \langle {\bf R}, {\bf R}\rangle = {\bf R}_{ijkl}
{\bf R}^{ijkl}, \ {\rm and} \ \langle {\bf R}, *{\bf R}\rangle = 
{\bf R}_{ijkl}(*{\bf R})^{ijkl},$$
vanish identically. The vacuum Einstein equations impose only the 
condition (1.5), i.e. that $h$ is harmonic as a function of $(x, y)$. 
Thus, the function $h$ may be an arbitrary function of $u$, and so is 
not controlled in any H\"older or Sobolev space. It is thus clear that 
there is no coordinate system in which a general metric {\bf g} of the 
form (1.4) is controlled in $L^{2,p}$, or even $C^{0}$.

\medskip

 To deal with this situation, one imposes bounds on the components of 
{\bf R} in a fixed coordinate system or framing. An efficient way 
to do this is to choose a future-directed unit time-like vector 
$T = e_{0}$ and extend it to an orthonormal frame $e_{\alpha}$, 
$0 \leq  \alpha \leq  n$, where the space-time dimension is $n+1.$ Since 
the space $T^{\perp}$ orthogonal to $T$ is space-like and $O(n)$ is 
compact, the particular choice of framing for $T^{\perp}$ is 
unimportant. The norm of {\bf R} w.r.t. $T$ is then defined as
\begin{equation} \label{e1.6}
|{\bf R}|_{T}^{2} = \sum (R_{ijkl})^{2}, 
\end{equation}
where the components are w.r.t. the framing $e_{\alpha}.$

 Observe that if, at a point $p\in{\bf M},$ the vector $T = T_{p}$ is 
contained in a compact subset $W$ of the future interior null cone 
$T_{p}^{+}{\bf M},$ then the norms (1.6) are all equivalent, with 
constant depending only on $W$. Hence, if $K$ is a compact subset of 
the space-time ({\bf M, g}) and $T$ is a continuous vector field on 
$K$, then $T$, (or more precisely, Im $T$, where $T$ is viewed as a 
section of the tangent bundle), lies within a compact subset of 
$T^{+}{\bf M}$, where $T^{+}{\bf M}$ is the bundle of future interior 
null cones in the tangent bundle $T{\bf M}$.

 To state the main result, we need the following definition, which is 
essentially just a normalization on the size of the region to be 
considered in ({\bf M, g}), as is (1.1).

\begin{definition} \label{d1.1}
{\rm  Let $\Omega $ be a domain in a smooth Lorentz manifold ({\bf M, g}). 
Then $\Omega $ satisfies the {\sf size conditions} if the following holds: 
The domain $\Omega$ admits a smooth time function $t$, with 
$c_{o}^{-1} \leq ||\nabla t|| \leq c_{o}$, for an arbitrary but fixed 
constant $c_{o} < \infty$. Further, one has 
\begin{equation} \label{e1.7}
C_{1} = B_{p}(1)\times [-1,1] \subset\subset  \Omega , 
\end{equation}
i.e. the 1-cylinder $C_{1}$ has compact closure strictly contained in 
$\Omega .$  Here $B_{p}(r)$ is the geodesic $r$-ball about a point $p$ in 
$S$, where $S = S_{0} = t^{-1}(0)$ and the metric $g$ on $S$ is that induced 
from {\bf g}. The product $B_{p}(1)\times [-1,1]$ is identified with a subset 
of $\Omega $ by the flow of $\nabla t,$ i.e. $(q, s) \rightarrow  
\gamma_{q}(s),$ where $\gamma_{q}(s)$ is the flow line of $\nabla t,$ 
starting at $q$ and terminating on the $s$-level set $S_{s} = t^{-1}(s)$. 

  Let $T = \nabla t/||\nabla t||$ be the corresponding future-directed unit 
time-like vector field, and set } 
\begin{equation} \label{e1.8}
D = {\rm Im} T|_{C_{1}} \subset\subset  T^{+}\Omega .
\end{equation}
\end{definition} 

 The size conditions represent a Lorentzian analogue of the condition 
(1.1). They can always be realized by choosing $\Omega $ to be a 
sufficiently small open set in ({\bf M, g}) and rescaling the metric up 
sufficiently. Essentially, they just serve to normalize the data.

 The main result of the paper is then the following:
\begin{theorem} \label{t 1.2.}
  Let $\Omega $ be a domain in a vacuum $(n+1)$-dimensional space-time 
({\bf M, g}), $n \geq 2$, satisfying the size conditions. Suppose that 
there are constants $C <  \infty $ and $v_{o} > $ 0 such that
\begin{equation} \label{e1.9}
|{\bf R}|_{T} \leq  C,  \ \ vol_{g}B_{p}(\tfrac{1}{2}) \geq  v_{o}. 
\end{equation}

 Then there exists a constant $r_{o} > 0$, depending only on $C$, 
$v_{o}$, $c_{o}$, $D$, (and $n$), and a coordinate system $(\tau , x_{i})$, 
$1 \leq  i \leq n$, on the $r_{o}$-cylinder 
\begin{equation} \label{e1.10}
C_{r_{o}} = D_{p}(r_{o})\times [- r_{o}, r_{o}] \subset  C_{1}, 
\end{equation}
such that the components of the metric ${\bf g}_{\alpha\beta}$ are in 
$C^{1,\gamma}\cap L^{2,p}$, for any $\gamma < 1$, $p < \infty$. Here 
$D_{p}(r)$ is the geodesic $r$-ball about $p$ in the level set 
$\tau = 0$ and the product structure is that induced by the flow of 
$\nabla \tau$. 

 Further, there exists a constant $R_{o} <  \infty ,$ depending only on 
$C$, $v_{o}$, $c_{o}$, $D$ and the exponent $p$, such that, on $C_{r_{o}},$
\begin{equation} \label{e1.11}
||{\bf g}_{\alpha\beta}||_{L^{2,p}} \leq  R_{o}. 
\end{equation}
More precisely, for any $k \leq  2$, and $0 \leq  \alpha ,\beta  \leq  
n$,
\begin{equation} \label{e1.12}
||\partial_{\mu}^{k}{\bf g}_{\alpha\beta}||_{L_{x}^{2-k,p}} \leq  
R_{o}, 
\end{equation}
where $\partial_{\mu}^{k}$ denotes any $k$-fold space-time partial 
derivative and the spatial $L_{x}^{2-k,p}$ norm is taken over any 
spatial slice $\{\tau = const\}$ in $C_{r_{o}}$. The constant $R_{o}$ 
is independent of $\tau $ in $[- r_{o}, r_{o}].$
\end{theorem}

 The coordinates in Theorem 1.2 are geometrically natural; the time 
coordinate $\tau$ is a Gaussian (equidistant) coordinate, 
while the spatial coordinates are chosen to be harmonic on the spatial 
slices $\{\tau = const\}$. It is not clear if there exist space-time 
harmonic, i.e wave, coordinates, in which {\bf g} has this degree of 
regularity.

\medskip

 The condition that ({\bf M, g}) is vacuum, i.e. ${\bf Ric_{g}} = 0$, 
is used in a rather minor way. It is only used, via the Bianchi 
identity, to obtain $L^{p}$ bounds on the $2^{\rm nd}$ time derivatives 
$\partial_{\tau}\partial_{\tau}{\bf g}_{0\alpha}$ of the components 
${\bf g}_{0\alpha}.$ This is equivalent to bounds on the $2^{\rm nd}$ 
time derivatives on the components of the shift vector of the 
coordinates, i.e. a bound on the acceleration of the shift. Such acceleration 
components do not appear in any component of the curvature tensor 
{\bf R}. All other estimates on $\partial_{\gamma}^{k}{\bf g}_{\alpha\beta}$ 
are independent of the Einstein equations.

 The vacuum condition can be weakened to an assumption on the 
stress-energy tensor ${\mathcal T} $ in the Einstein equations, of the 
form
\begin{equation} \label{e1.13}
||\nabla_{T}{\mathcal T}||_{L_{x}^{-1,p}} \leq  C. 
\end{equation}
Here $L_{x}^{-1,p}$ is the dual space of $L_{o}^{1,q},$ the space of 
$L_{x}^{1,q}$ functions of compact support on spatial slices 
$\{\tau  = const\}$ within $C_{r_{o}}$, $p^{-1}+q^{-1} = 1$. 

  The condition (1.13) will be satisfied automatically for many physically 
natural matter fields, cf. Remark 2.1.

\medskip

 Theorem 1.2 is formulated in such a way that it is easy to pass to 
limits. Thus, suppose $({\bf M}_{i}, {\bf g}_{i})$ is a sequence of 
smooth space-times satisfying the hypotheses of the Theorem. There 
exist then domains $\Omega_{i} \subset  ({\bf M}_{i}, {\bf g}_{i}),$ 
points $p_{i}\in\Omega_{i}$ such that the size conditions (1.7)-(1.8) 
hold, with $D_{i}$ uniformly compact in $T^{+}\Omega_{i},$ (i.e. $T$ 
does not become arbitrarily close to null cones). If (1.9) and (1.13) 
hold uniformly on $\Omega_{i},$ then there is a subsequence which 
converges to a limit $C^{1,\gamma}\cap L^{2,p}$ space-time ({\bf M, g}), 
defined at least on an $r_{o}$-cylinder $C_{r_{o}}$. Further, the 
convergence to the limit is $C^{1,\gamma}$ and weak $L^{2,p},$ and the 
bound (1.12) holds on the limit.

\medskip

 Define a Lorentz manifold  ({\bf M, g}) to be {\sf weakly regular} if 
{\bf g} is a continuous Lorentz metric, with 
${\bf g}\in L^{1,2}_{loc}({\bf M})$. It is well-known, cf. [11], [9] 
for example, that such metrics have a well-defined curvature tensor 
{\bf R} in the sense of distributions. This leads to the following corollary.

\begin{corollary} \label{c 1.3.}
  Let ({\bf M, g}) be a weakly regular Lorentz manifold, and let 
$\Omega \subset \subset {\bf M}$ be a domain with compact closure in 
{\bf M}. Suppose the size conditions hold locally on $\Omega$, in that 
the constant $\frac{1}{2}$ is replaced by a small constant $\delta_{o}$ so 
that $B_{p}(\delta_{o}) \subset \subset {\bf M}$, for any $p \in \Omega$. 
Suppose also the bounds (1.9) hold locally and uniformly on $\delta_{o}$ 
cylinders as in (1.7), centered at any $p \in \Omega$. 

  Then $\Omega$ may be covered by a finite atlas of charts in which the 
metric ${\bf g} = {\bf g}_{\alpha\beta}$ satisfies all the bounds in (1.12), 
except for the $L_{x}^{p}$ bound on 
$\partial_{\tau}\partial_{\tau}{\bf g}_{0\alpha}$. The bounds in (1.12) 
depend, near $\partial \Omega$, on the distance of $\partial \Omega$ to 
$\partial {\bf M}$.

 If in addition the bound (1.13) holds distributionally on {\bf M}, (e.g. 
({\bf M, g}) is a weak solution of the vacuum equations), then all bounds 
in (1.12) hold locally on $(\Omega, {\bf g})$. 
\end{corollary}

 We refer to the proof of Corollary 1.3 below for the precise meaning 
that (1.13) holds distributionally.

\medskip

 The proof of Theorem 1.2 and Corollary 1.3 follow in \S 2, while \S 3 
concludes the paper with several remarks and extensions of these 
results, together with some open problems.

\section{Proofs of the Results.}

\setcounter{equation}{0}

 In this section, we prove Theorem 1.2 and Corollary 1.3. For clarity, 
the proof of Theorem 1.2 is divided into several steps, each treating 
basically separate issues. In the following, as already above, 
space-time quantities are generally denoted in boldface while spatial 
quantities are not in boldface.

\medskip

{\bf Step I. (Initial Choice of Domain).}

 Let $B_{p}(r)$ be the intrinsic geodesic ball about $p$ in $S$. Since $S$ 
is achronal in the cylinder $C_{1}$ from (1.7), the extrinsic radius of 
$B_{p}(r)$ is bounded below for $r$ small. Thus, if $\gamma$ is any 
space-like curve in ({\bf M, g}) from $p$ to 
$x \in \partial B_{p}(r_{1}) \subset S$, for $r_{1}$ small, then the length 
$L(\gamma)$ satisfies $L(\gamma) \geq l_{o}r_{1}$; the constants $l_{o}$ and 
$r_{1}$ depend only on $c_{o}$ and $D$ in Definition 1.1. 

  Let ${\mathcal D}_{r_{1}}$ be the domain of dependence of $B_{p}(r_{1})$ 
in the manifold ({\bf M, g}). Thus, by choosing $r_{1}$ sufficiently small, 
again depending only on $c_{o}$, $D$, one has
\begin{equation} \label{e2.1}
{\mathcal D}_{r_{1}}\subset\subset  \Omega, 
\end{equation}
i.e. ${\mathcal D}_{r_{1}}$ has compact closure in $\Omega$.

 The region ${\mathcal D}_{r_{1}}$ is globally hyperbolic and hence any 
pair of time-related points in ${\mathcal D}_{r_{1}}$ may be  joined by 
a time-like maximizing geodesic in ${\mathcal D}_{r_{1}}.$ Recall from 
Definition 1.1 that the curves $\gamma_{x}$ are the flow lines of $\nabla t$ 
through $x$. For $r_{2} > $ 0 small, (to be determined below), let 
\begin{equation} \label{e2.2}
q = \gamma_{p}(- r_{2}), 
\end{equation}
so $q << p$, i.e. $q$ is to the past of $p$. For $x$ to the future of $p$, 
$x >> p$, let
\begin{equation} \label{e2.3}
\tau (x) = dist_{{\bf g}}(x, q) -  dist_{{\bf g}}(p, q), 
\end{equation}
so that $\tau (p) = 0$ and $\tau(x) > 0$, for $x >> p$. The distance 
$\tau (x)$ is realized by a maximizing time-like geodesic 
$\sigma_{v}(\tau) = exp_{q}(\tau + \tau_{o})v$ from $q$ to $x$; here $v 
\in T_{q}^{+}{\bf M}$, with ${\bf g}(v,v) = -1$, and $\tau_{o} = 
dist_{\bf g}(p, q)$. This normalization gives $p = \sigma_{v_{o}}(0)$, 
for some $v_{o} \in T_{q}^{+}{\bf M}$. Let 
$$N = \nabla\tau $$
be the corresponding unit time-like vector field, so that $N$ is the 
tangent vector to geodesics $\sigma$ issueing from $q$. Although $N$ is 
well-defined and smooth along the individual geodesics 
$\sigma_{v}(\tau)$, (for any $\tau$ until one reaches the boundary of 
({\bf M, g})), at points where the exponential map $exp_{q}$ has cut or 
conjugate points, $N$ is not uniquely defined. Of course, past such cut 
or conjugate points, the geodesics $\sigma_{v}(\tau)$ are no longer 
maximal. Thus, in general, $\tau$ is merely Lipschitz and $N$, as a 
vector field, is defined only almost everywhere to the future of $q$. 
At the end of Step I, (via the work in Step V), it will be seen that in fact 
$\tau$ and $N$ are smooth, in suitable domains of a definite size. In the 
following, unless stated otherwise, all geodesics are assumed to be maximal, 
i.e. they are not continued past the conjugate or cut points of $q$.

 Next, let $\Sigma  = \Sigma_{0} = \tau^{-1}(0)$ and similarly let 
$\Sigma_{\tau}$ be the $\tau$-level set of $\tau$ in ${\mathcal 
D}_{r_{1}}$. Since the geodesics are maximal, $\sigma_{v}(\tau) \in 
\Sigma_{\tau}$, and $p \in \Sigma = \Sigma_{0}$. Again, in general, 
$\Sigma_{\tau}$ is only Lipschitz. For $r_{2} \leq r_{1}$, consider the 
intrinsic geodesic ball $B_{p}(r_{2}) \subset S$ and let 
\begin{equation} \label{e2.4}
{\mathcal C} = \{x\in {\mathcal D}_{r_{1}}: x = \sigma (\tau ), \ \tau 
\leq  r_{2}, \ {\rm and} \ \sigma (\tau) \cap  B_{p}(r_{2}) \neq 
\emptyset \} \subset {\mathcal D}_{r_{1}}.
\end{equation}
This is the ``cone'' of maximal geodesics $\sigma $ starting at $q$, 
hitting $S $ within $B_{p}(r_{2}),$ and terminating at time $\tau  = 
r_{2}.$ 

 Observe that the vector field $N$ restricted to ${\mathcal C}$ stays 
within a compact subset of $T^{+}{\bf M}$. In fact, since $N$ is 
parallel along its geodesic flow lines, this needs to be verified only 
at the base point $q$, where it holds by construction. It then follows 
from the curvature bound (1.9) and the remarks following (1.6) that
\begin{equation} \label{e2.5}
|{\bf R}|_{N} \leq  C_{1} = C_{1}(C, D). 
\end{equation}

 Now the curvature bound (2.5) and the Rauch comparison theorem, cf. 
[2] for instance, imply that if $r_{2}$ is sufficiently small, 
depending (explicitly) only on $C_{1},$ then the exponential map 
$exp_{q}$ restricted to the interior future null cone in $T_{q}^{+},$ 
has no conjugate points in ${\mathcal C}.$ Thus, $exp_{q}$ is of 
maximal rank, and so a local diffeomorphism on ${\mathcal C}.$ In fact, 
for $r_{1}$ is sufficiently small, again depending only on $C$ and 
$c_{o}$, no time-like or null geodesic within ${\mathcal D}_{r_{1}}$ has 
conjugate points, and so $exp_{x}$ is of maximal rank on time-like 
geodesics in ${\mathcal D}_{r_{1}}$, $\forall x \in {\mathcal 
D}_{r_{1}}$.

 Since ${\mathcal D}_{r_{1}}$ is globally hyperbolic and without 
time-like conjugate points, it follows then from [2, Thm.11.16] for 
instance, that any pair of points $x, y \in {\mathcal D}_{r_{1}}$ with 
$y >> x$, may be joined by a {\it unique} maximizing time-like geodesic 
in ${\mathcal D}_{r_{1}}$, provided ${\mathcal D}_{r_{1}}$ is future 
1-connected, (i.e. any pair of time-like curves joining $x$ and $y$ are 
homotopic through time-like curves).

\medskip

  In general, ${\mathcal D}_{r_{1}}$ need not be future 1-connected. 
Consider for example the past null cone of 2-dimensional Minkowski 
space in hyperbolic coordinates $t \in (-\infty, 0)$, $\phi \in 
(-\infty, \infty)$, 
$$-dt^{2} + t^{2}d\phi^{2}.$$
If $\phi$ is identified periodically, with any period, then the 
resulting space-time is globally hyperbolic but not future 1-connected. 
The future exponential map based at any point $q$ has cut points; if 
the period of $\phi$ is sufficiently small, or if $q$ is sufficiently 
close to $\{0\}$, then cut points occur arbitrarily close to $q$.

  However, ${\mathcal D}_{r_{1}}$ is future 1-connected if it has a 
simply connected Cauchy surface $S$, i.e. $S \equiv S_{r_{1}} = S \cap 
{\mathcal D}_{r_{1}}$, for $S$ as in (1.7ff). To see this, let 
$\gamma_{1}$, $\gamma_{2}$ be two time-like curves with common 
endpoints in ${\mathcal D}_{r_{1}}$. The flow of the time-like vector 
field $\nabla t$ gives a strong deformation retraction of $\gamma_{1} 
\cup \gamma_{2}$ onto a closed loop $\lambda_{1} \cup \lambda_{2}$ in 
$S$. If $S$ is simply connected, then $\lambda_{1}$ may be deformed 
into $\lambda_{2}$ within $S$. These two homotopies, time-like along 
$\nabla t$ and space-like along the Cauchy surfaces $S_{t}$ may be 
performed simultaneously, but with the latter at a larger speed than 
the former, to produce a time-like homotopy from $\gamma_{1}$ to 
$\gamma_{2}$.

\medskip

  We will prove later that $S$, (or more precisely, a domain in $S$ of 
a definite size), is simply connected. However, in order not to 
overburden the arguments to follow with such further issues, we assume 
in the following, through Step IV, that the Cauchy surface $S \subset 
{\mathcal D}_{r_{1}}$ is simply connected. This hypothesis will be 
removed in Step V, using the results obtained in the previous steps.

\medskip

  It follows then that the exponential map $exp_{q}$ is a 
diffeomorphism onto ${\mathcal C}$, when restricted to a suitable domain 
in $T_{q}^{+}{\bf M}$. The time function $\tau$ is smooth in 
${\mathcal C} \setminus \{q\}$, as are the level sets $\Sigma_{\tau} 
\cap {\mathcal C}$, and there is a unique maximizing geodesic from $q$ 
to any point in ${\mathcal C}$.

  From now on, we consider $\Sigma_{\tau} \subset {\mathcal C}$, and so 
let $\Sigma_{\tau}$ denote the prior $\Sigma_{\tau}$ intersected with 
${\mathcal C}$. The level sets $\Sigma_{\tau}$ form a foliation of 
${\mathcal C}$ by equidistant space-like hypersurfaces, with unit 
normal $N$.

\medskip

{\bf Step II. (Initial Curvature and Volume Estimates).}

 The geodesic congruence $\sigma $ on $({\mathcal C}, {\bf g})$ 
satisfies the Riccati or transport equation
\begin{equation} \label{e2.6}
K'  + K^{2} + {\bf R}_{N} = 0. 
\end{equation}
Here $K = D^{2}\tau $ is the $2^{\rm nd}$ fundamental form or extrinsic 
curvature of the leaves $\Sigma_{\tau}$, ${\bf R}_{N}$ is the symmetric 
bilinear form given by ${\bf R}_{N}(X) = \langle {\bf R}(N,X)X, N 
\rangle$ and $' $ is the covariant derivative in the direction $N$. 
Hence, the bound (2.5) gives
$$|K'  + K^{2}| \leq  C_{1}. $$
This estimate holds on $\Sigma_{\tau}$, for all $\tau \in [-r_{2}, 
r_{2}]$. It then follows by standard comparison theory for the Riccati 
ODE (2.6) that if $r_{2}$ is sufficiently small, depending only on $c_{o}$, 
$D$, and $C_{1}$, then
\begin{equation} \label{e2.7}
|K|_{L^{\infty}} \leq  C_{2},  \ \ |K'|_{L^{\infty}} \leq  C_{2}, 
\end{equation}
on all $\Sigma_{\tau}$, $\tau \in [-\frac{r_{2}}{2}, r_{2}]$. The constant 
$C_{2}$ depends only on $r_{2}$ and $C_{1}$. The Gauss equation 
relating the curvature {\bf R} of the ambient manifold ({\bf M, g}) 
with that of the spatial slices $\Sigma_{\tau}$ reads
\begin{equation} \label{e2.8}
{\bf R}_{ijkl} = R_{ijkl} + K_{ik}K_{jl} -  K_{il}K_{jk}, 
\end{equation}
for spatial components $(ijkl)$. This, together with the bounds (2.5) 
and (2.7), thus gives the bound
\begin{equation} \label{e2.9}
|R_{g}|_{L^{\infty}} \leq  C_{3}, 
\end{equation}
on $\Sigma_{\tau}$, $\tau \in [-\frac{r_{2}}{2}, r_{2}]$. Let $dK$ be 
the exterior derivative, (w.r.t. the connection induced by $g$), of 
$K$, when $K$ is viewed as a 1-form with values in $T\Sigma_{\tau}$, 
i.e. $dK(X,Y,Z) = (\nabla_{X}K)(Y,Z) - (\nabla_{Y}K)(X,Z)$. The 
Gauss-Codazzi equations are
$$dK = {\bf R}^{N}, $$
where ${\bf R}^{N}(X,Y,Z) = \langle {\bf R}(N,X),Y,Z \rangle$. Hence, 
(2.5) also implies
\begin{equation} \label{e2.10}
|dK|_{L^{\infty}} \leq  C_{3}. 
\end{equation}

 We record also the well-known constraint equations:
\begin{equation} \label{e2.11}
\delta K = - dH - {\bf Ric}(N), 
\end{equation}
$$R -  |K|^{2} + H^{2} = 2{\bf Ric}(N,N) + {\bf R}, $$
where $H =$ tr $K$ is the mean curvature and the operators $\delta$ and 
$d$ are taken on $\Sigma_{\tau}$.

\medskip

 Next, we use the bounds above to obtain a lower volume bound on the 
spatial slices $\Sigma_{\tau}$, $\tau \in [-\frac{r_{2}}{2}, r_{2}]$, 
from that on the slice $S = S_{0}$ in (1.9). To do this, let $\widehat 
S = S \cap {\mathcal C}$. The domain $\widehat S$ may be written as a 
graph over $\Sigma = \Sigma_{0} \subset {\mathcal C}$ via the time 
coordinate $\tau$ in the usual way. Thus, each geodesic $\sigma  = 
\sigma_{v}$ intersects $\Sigma$ and $\widehat S$ in exactly two points 
$\sigma(\tau_{1})$, $\sigma(\tau_{2})$, with $|\tau_{i}| \leq  r_{2}$. 
For $x = \sigma_{v}(\tau_{1})\in \Sigma$, let $u(x) = \tau_{2}- 
\tau_{1}$, so that $\sigma_{v}(\tau_{2})\in\widehat S.$ This gives a 
diffeomorphism $\phi : \Sigma  \rightarrow  \widehat S,$ and hence
$$vol\widehat S = \int_{\widehat S}dV_{\widehat S} = 
\int_{\Sigma}\phi^{*}(dV_{\widehat S}) = \int_{\Sigma}JdV_{\Sigma}, $$
where $J = detD\phi $ is the Jacobian of $\phi .$ Since both $\widehat 
S$ and $\Sigma $ are space-like, the function $u$ is a Lipschitz 
function, (cf. [11]), whose (weak) derivative is uniformly bounded, 
since both normal vectors $N$ and $T$ lie in compact subsets of 
interior null cones. In addition, the (uniform) time $\tau $ exponential 
map, mapping $\Sigma$ to $\Sigma_{\tau}$ has Jacobian uniformly bounded 
above and below on $[-\frac{r_{2}}{2}, r_{2}]$, by the bound (2.7). 
(Recall that $H$ = tr$K$ measures the infinitesimal volume expansion or 
contraction). It follows that the Jacobian $J$ is uniformly bounded 
below, (depending only on $C$, $D$). Hence, 
\begin{equation} \label{e2.12}
vol\Sigma  \geq  v_{1}\cdot  vol\widehat S.
\end{equation}
Now the lower bound on $volB_{p}(\frac{1}{2}) \subset  S$ in (1.9) does 
not immediately imply a lower bound on $vol\widehat S$; (it could 
apriori happen that most all of the volume of $B_{p}(\frac{1}{2})$ 
occurs outside $\widehat S$). However, in this case one can repeat all 
the estimates (2.6)-(2.9) when the construction of ${\mathcal C}$ is 
based at other center points $q' $ in place of $q$. Thus, for $p'\in 
B_{p}(\frac{1}{2}) \subset S$, define $q'$ as in (2.2) and let 
${\mathcal C}'$ be then as in (2.4). The same estimates as above then hold 
in ${\mathcal C}'$. The corresponding domains $\widehat S'  = S\cap 
{\mathcal C}'$ give a covering of $B_{p}(\frac{1}{2}) \subset S$. Hence 
the volume bound in (1.9) and the estimates above now do give the 
existence of points $p_{o}\in B_{p}(\frac{1}{2}) \subset S$ such that
\begin{equation} \label{e2.13}
vol\Sigma_{p_{o}} \geq  v_{2} > 0, 
\end{equation}
where $\Sigma_{p_{o}}$ is the level set of $\tau$, (i.e. $\tau_{o}$), 
containing $p_{o}$ and $v_{2} = v_{2}(v_{o}, D, C_{1})$. (The local estimate 
(2.12) does not in fact depend on the absence of future cut points of 
$exp_{q}$, cf. the discussion concerning (2.59) below).

 Recall the standard volume comparison theorem in Riemannian geometry: 
if $(N, g)$ is a Riemannian $n$-manifold, with $Ric_{g} \geq  - 
(n-1)k,$ then the ratio
$$\frac{vol D(r)}{vol D_{k}(r)}, $$
is monotone non-increasing in $r$. Here $D(r)$ denotes the volume of a 
geodesic $r$-ball at any fixed point, while $D_{k}(r)$ is the geodesic 
$r$-ball in the $n$-dimensional space form of constant curvature $k$. 
It then follows from the curvature bound (2.9), together with (2.13), 
that the geodesic balls $D_{p_{o}}(r) \subset  (\Sigma_{p_{o}}, g)$ satisfy
$$volD_{p_{o}}(r) \geq  v_{3}r^{n},$$ 
for all $r \leq  r_{2},$ where $v_{3}$ depends only on $v_{2}$ and 
$C_{3}$. 

 Observe that a similar estimate also holds for geodesic balls on other 
spatial slices $\Sigma_{\tau}$, with $\Sigma_{0} = \Sigma_{p_{o}}$, for 
$\tau \in [-\frac{r_{2}}{2}, r_{2}]$. Namely, the $L^{\infty}$ bound on $K$ 
in (2.7) bounds the infinitesimal distortion in the spatial metrics, and 
hence distances and volumes, under the flow of $N$. It follows that within 
the cylinder ${\mathcal C}_{o}$ centered at $p_{o}$, the volume estimate 
above holds for balls $D_{p_{\tau}}(r) \subset \Sigma_{\tau}$; thus, for 
$p_{\tau} = \sigma_{v_{o}}(\tau)$, where $p_{o} = \sigma_{v_{o}}(0)$, and 
for $r \leq r_{2}$, one has
\begin{equation} \label{e2.14}
volD_{p_{\tau}}(r) \geq  v_{4}r^{n}, 
\end{equation}
$v_{4} = v_{4}(v_{o}, C, D)$. An upper bound on the volume of 
$volD_{p_{\tau}}(r)$ of the form (2.14) follows immediately from the 
curvature bound (2.9). 

  In the construction above, we have shifted the original base point 
$p$ to a new base point $p_{o}$. However, one may now use these 
estimates to obtain equivalent volume bounds for the slices 
$\Sigma_{\tau}$ within the original cylinder ${\mathcal C}$ centered at $p$. 
This may be done by constructing a suitable chain, of bounded cardinality, 
of overlapping cylinders ${\mathcal C}_{i}$ from ${\mathcal C}_{o}$ to 
${\mathcal C}$. One then uses the arguments above on each ${\mathcal C}_{i}$, 
together with the fact that upper and lower volume bounds of spatial slices 
are equivalent to upper and lower volume bounds of each cylinder 
${\mathcal C}_{i}$.

  Thus, in the following, we work on the original cylinder ${\mathcal 
C}$ from (2.4) centered at $p$; the bound (2.14) holds with $p$ in 
place of $p_{o}$.

\medskip

{\bf Step III. (Local Coordinates).}

 In this step, we define the cylinder $C_{r_{o}}$ and the local 
coordinate system on it, and obtain in addition some initial estimates 
on ${\bf g}_{\alpha\beta}.$ The local coordinates are 
Gaussian in time and harmonic in space, (Gaussian-harmonic coordinate 
system). 

  Thus, the function $\tau$ from (2.3) is chosen as the time coordinate 
on ${\mathcal C}$. To construct spatial harmonic coordinates, start 
with the slice $\Sigma  = \Sigma_{0}$ within ${\mathcal C}$. By (2.9) 
and (2.14), one has the bounds
\begin{equation} \label{e2.15}
|R_{g}| \leq  C_{3}, \ \ vol_{g}D_{p}(r_{2}) \geq  v_{4}. 
\end{equation}
It then follows, for instance from the discussion in \S 1, that there 
exists $r_{o} >$ 0, depending only on $C_{3}$ and $v_{4},$ such that 
the geodesic ball $D_{p}(r_{o}) \subset \Sigma $ admits a harmonic 
coordinate system $\{x_{i}\}$, $1 \leq  i \leq n$, in which the spatial 
metric $g = {\bf g}|_{\Sigma}$ is controlled in $L^{2,p}$, i.e.
\begin{equation} \label{e2.16}
||g_{ij}||_{L^{2,p}} \leq  R_{o}, 
\end{equation}
where the $L^{2,p}$ norm is taken on $D_{p}(r_{o})$, and 
$R_{o} = R_{o}(C_{3}, v_{4}, p)$. The harmonic functions $x_{i}$ are 
solutions to the Dirichlet problem
\begin{equation} \label{e2.17}
\Delta_{g}x_{i} = 0, \ \ x_{i}|_{\partial D} = \phi_{i}, 
\end{equation}
where $D = D_{p}(r_{o})$ and $\phi_{i}$ are suitably chosen boundary 
values, (approximating linear-type functions, cf. [12]).

 Let $\phi_{i,\tau} = \phi_{i}\circ\psi_{\tau},$ where $\psi_{\tau}$ is 
the time $\tau $ flow from $\Sigma_{\tau}$ to $\Sigma_{0}$ along the 
integral curves of $N$. Thus, $\psi_{\tau}$ maps a domain $D_{\tau} 
\subset \Sigma_{\tau}$ diffeomorphically onto $D$ and $\phi_{i,\tau}$ 
are functions defined on $\partial D_{\tau}.$ It follows that
\begin{equation} \label{e2.18}
N(\phi_{i,\tau}) = 1 \ {\rm at} \  \partial D_{\tau}.  
\end{equation}
Define the functions $x_{i}$ on $D_{\tau}$ to be solutions to the 
Dirichlet problem
\begin{equation} \label{e2.19}
\Delta_{g_{\tau}}x_{i} = 0, \ \ x_{i}|_{\partial D_{\tau}} = 
\phi_{i,\tau}. 
\end{equation}

 By (2.9) and (2.14), the estimate (2.15) holds uniformly on 
$D_{\tau},$ for $\tau \in [-\frac{r_{2}}{2}, r_{2}]$. Hence, as with 
$\Sigma_{0}$, $r_{o} > 0$ may be chosen, depending only on $C_{3}$ and 
$v_{4},$ such that the functions $\{x_{i}\}$ form a harmonic coordinate 
system on $D_{\tau} \subset \Sigma_{\tau}$, on which one has the bounds
\begin{equation} \label{e2.20}
||g_{ij}||_{L^{2,p}} \leq  R_{o}, 
\end{equation}
where the $L^{2,p}$ norm is taken on $D_{\tau}$ and $g = g_{\tau}.$ The 
estimate (2.20) holds for all $|\tau| \leq  r_{o}.$

\medskip

 This construction gives the local coordinate system $(\tau, x_{i})$, 
$1 \leq i \leq  n$, on the $r_{o}$-cylinder 
\begin{equation} \label{e2.21}
C_{r_{o}} = D_{p}(r_{o})\times [- r_{o}, r_{o}]
\end{equation}
about $p$, where the product structure is defined by the 
flow of $\nabla \tau$. For the remainder of the proof, $\Sigma_{\tau}$ 
is now redefined to be its intersection with $C_{r_{o}}$, i.e. 
$\Sigma_{\tau} \equiv D_{\tau}$. 

\medskip

  The metric {\bf g} in these coordinates has the form
\begin{equation} \label{e2.22}
{\bf g} = (- 1 + |\xi|^{2})(d\tau )^{2} + g_{ij}(dx_{i} + \xi_{i}d\tau 
)(dx_{j} + \xi_{j}d\tau ), 
\end{equation}
where $\xi  = \{\xi_{i}\}$ is the shift vector. Thus,
$$\partial /\partial\tau  = N + \xi , $$
with $N = \nabla\tau$. The lapse function $\alpha$ of this foliation 
satisfies $\alpha  \equiv 1$.

 On each slice $\Sigma_{\tau} \subset C_{r_{o}}$, one has good spatial 
control, namely for $g_{ij} = {\bf g}_{ij} = {\bf g}|_{\Sigma{\tau}}$, 
(2.20) holds. As usual, Latin indices $i,j$, denote spatial variables, 
i.e. $1 \leq i,j \leq n$, while Greek indices $\alpha, \beta$ denote 
space-time variables, $0 \leq \alpha ,\beta  \leq n$.

 In the following, all Sobolev norms $L^{k,p}$ are understood to be 
spatial norms, i.e. the derivatives and norms are taken on spatial 
leaves $\Sigma_{\tau}$. Thus, for emphasis or clarity, we sometimes 
write $L_{x}^{k,p}$ in place of $L^{k,p}$. All estimates will be 
independent of $\tau$, for $\tau \leq r_{o}$.

\medskip

{\bf Step IV. $(L^{2,p}$ Estimates of $g_{\alpha\beta})$.}

 In this next step, we extend the estimate (2.20) to include the 
remaining terms ${\bf g}_{0\alpha}$, $0 \leq  \alpha  \leq  n$, and 
also obtain estimates on the time derivatives of ${\bf 
g}_{\alpha\beta}.$

 Before beginning, we first improve the estimate (2.7) on the $2^{\rm 
nd}$ fundamental form. Recall the Simons', (or Bochner-Weitzenbock) 
formula, cf. [3, Ch. 1I]:
$$D^{*}DK = \delta dK + d\delta K - {\mathcal R} (K),$$ 
on $(\Sigma_{\tau}, g_{\tau}),$ where the term ${\mathcal R} (K)$ is 
linear in the curvature and $K$; the exact form of ${\mathcal R} (K)$ 
plays no role in the argument, but for completeness is given by 
${\mathcal R} (K) = Ric\circ K + K\circ Ric -  2R\circ K,$ where 
$R\circ K$ is the action of the curvature tensor $R$ on symmetric 
bilinear forms. The elliptic operator $D^{*}D = - trD^{2}$ is the 
so-called rough Laplacian.

 In the following, we frequently write $f\in L^{k,p}$ or $f \in 
L_{x}^{k,p}$ as shorthand for $f$ is uniformly bounded in $L^{k,p}$ 
along the spatial slices $\Sigma_{\tau}$, $|\tau| \leq r_{o}$.

 By (2.10), $dK \in  L^{\infty},$ and hence $\delta dK \in  
L_{x}^{-1,p},$ for all $p <  \infty ;$ recall that these spaces are 
defined as following (1.13). Similarly, by (2.11), since $dd = 0$,
\begin{equation} \label{e2.23}
d\delta K = d({\bf Ric}N) \in  L_{x}^{-1,p};
\end{equation}
here we recall that the operators $\delta$ and $d$ are spatial. The 
term ${\mathcal R}(K)$ is also bounded in $L^{\infty}.$ 
Hence, one has
\begin{equation} \label{e2.24}
D^{*}DK = Q_{1},
\end{equation}
where $Q_{1}$ is uniformly bounded in $L_{x}^{-1,p},$ for any $p <  
\infty$, while $K$ is uniformly bounded in $L^{\infty}$. By (2.20), the 
coefficients of $D^{*}D$ in the local coordinates $\{x_{i}\}$ are 
controlled in $L^{2,p} \subset C^{1,\gamma}$. It then follows from 
standard elliptic regularity theory, cf. [10], that
\begin{equation} \label{e2.25}
|K|_{L_{x}^{1,p}} \leq  C_{2}, 
\end{equation}
where $C_{2} = C_{2}(C, v_{o}, c_{o}, D, p)$, on all spatial slices 
$\Sigma_{\tau}$, $|\tau| \leq r_{o}$.

\medskip

{\bf Spatial Estimates.}

 Here, we prove that the components ${\bf g}_{0\alpha}$ also satisfy 
the $L^{2,p} = L_{x}^{2,p}$ estimate (2.20) uniformly on 
$\Sigma_{\tau}.$ One has $N = {\bf g}^{0\alpha}\partial_{\alpha}, 
\nabla x_{i} = {\bf g}^{i\alpha}\partial_{\alpha}.$ Hence, $N(x_{i}) = 
\langle N, \nabla x_{i} \rangle = {\bf g}^{0\alpha}{\bf g}^{i\beta}{\bf 
g}_{\alpha\beta} = {\bf g}^{0i}.$

\medskip

 To obtain estimates on $N(x_{i}),$ differentiate the harmonic 
coordinate condition (2.17), in the normal, (i.e. $N$), direction. Let 
$x_{i}'  = N(x_{i}).$ Since $\Delta x_{i} =$ 0, a standard computation, 
cf. [3, 1.184] for example, gives
\begin{equation} \label{e2.26}
\Delta x_{i}'  = -\Delta' x_{i} = \langle D^{2}x_{i}, K \rangle  - 
\langle dx_{i}, \delta K + \tfrac{1}{2}dH \rangle. 
\end{equation}
Here, as above and in the following, all metric quantities in (2.26) 
are on spatial slices $\Sigma_{\tau}.$

 By (2.25), $K$ is uniformly bounded in $L^{1,p}.$ The term 
$D^{2}x_{i}$ is also uniformly bounded in $L^{1,p},$ since by (2.20) 
the spatial metric is uniformly bounded in $L^{2,p}$ and hence the 
coordinate functions are uniformly bounded in $L^{3,p}.$ Further, both 
$\delta K$ and $dH$ are uniformly bounded in $L^{p}.$ Thus, 
$$\Delta x_{i}'  = Q_{2},$$
where $Q_{2}$ is uniformly bounded in $L^{p}, |\tau| \leq  r_{o}.$ As 
before, the coefficients of $\Delta $ are controlled in $L^{2,p}.$ 
Further, by construction, cf. (2.18), $x_{i}'  =$ 1 on 
$\partial\Sigma_{\tau}.$ Hence, standard elliptic regularity again gives
\begin{equation} \label{e2.27}
||x_{i}'||_{L^{2,p}} = ||{\bf g}^{0i}||_{L^{2,p}} \leq  C_{4}, 
\end{equation}
where $C_{4} = C_{4}(C, v_{o}, c_{o}, D, p)$. Observe also that 
\begin{equation} \label{e2.28}
{\bf g}^{00} = -1, 
\end{equation}
(since the lapse function $\alpha  \equiv 1$). Hence 
${\bf g}^{0\alpha} \in L^{2,p}$, i.e. the $L^{2,p}$ norm of 
${\bf g}^{0\alpha}$ is uniformly bounded, $0 \leq  \alpha  \leq  n$. 

 From this and (2.20), it is then an elementary exercise in linear 
algebra to see that 
\begin{equation} \label{e2.29}
||{\bf g}_{\alpha\beta}||_{L^{2,p}} \leq  C_{5}. 
\end{equation}
Briefly, ${\bf g}^{0\gamma} = (det{\bf 
g}_{\alpha\beta})^{-1}A_{0\gamma},$ where $A_{0\gamma}$ is the 
$(0,\gamma )$ cofactor in the matrix ${\bf g}_{\alpha\beta}.$ The 
cofactor $A_{00}$ involves only $g_{ij},$ and hence by (2.20), 
$A_{00}\in L^{2,p}.$ Thus $det{\bf g}_{\alpha\beta} \in  L^{2,p}.$ The 
same reasoning on ${\bf g}^{0\alpha}$ then gives $A_{0\alpha}\in 
L^{2,p},$ for all $\alpha .$ Each determinant $A_{0k}$ may be expanded 
along the first column to obtain a linear form in the variables ${\bf 
g}_{0i},$ with coefficients $(n-1)\times (n-1)$ determinants. Thus, one 
has a linear system of $n$ equations in $n$ unknowns ${\bf g}_{0i}.$ 
The matrix of this system is the $(n-1)$-compound $G_{n-1}$ of the 
matrix $[g_{ij}],$ i.e. $(G_{n-1})_{kl} = detA_{kl},$ where $A_{kl}$ is 
the $(k,l)$ cofactor of $[g_{ij}].$ Since $[g_{ij}]$ is non-singular, 
and since $[g_{ij}]$ non-singular implies that $G_{n-1}$ is 
non-singular, (cf. [8, \S 1.4] for instance), it follows that this 
linear system is invertible.

 The components ${\bf g}_{0i}$ are rational expressions in $\{g_{ij}\}$ 
and $\{A_{0k}\},$ each of which is now bounded in $L^{2,p}.$ Hence 
${\bf g}_{0i}$ is bounded in $L^{2,p}.$ Finally, since 1 $= {\bf 
g}^{0\alpha}{\bf g}_{\alpha 0} = {\bf g}^{00}{\bf g}_{00} + {\bf 
g}^{0i}{\bf g}_{i0},$ it follows from (2.28) that ${\bf g}_{00}$ is 
also bounded in $L^{2,p}.$ This establishes the bound (2.29).

\medskip

 Recall that $\partial_{\tau} = N + \xi ,$ while $\langle N, 
\partial_{i} \rangle = 0$, $i > 0$, by construction. Since $\langle 
\partial_{i}, \partial_{\tau}\rangle = {\bf g}_{i0}$ is bounded in 
$L^{2,p},$ it follows that $\langle \xi , \partial_{i} \rangle$ is 
bounded in $L^{2,p}.$ Hence the shift vector $\xi  = \{\xi_{i}\}$ is 
bounded in $L^{2,p},$ 
\begin{equation} \label{e2.30}
||\xi||_{L^{2,p}} \leq  C_{6}. 
\end{equation}
This completes the $L^{2,p}$ estimates of ${\bf g}_{\alpha\beta}$ in 
spatial directions.

\medskip

{\bf $1^{\rm st}$ Time Derivatives.}

 Next we turn to estimates on the time derivatives of ${\bf 
g}_{\alpha\beta},$ i.e. $L^{1,p}$ estimates for $\partial_{\tau}{\bf 
g}_{\alpha\beta}$. To begin, using the Leibniz rule, and the fact that 
$[\partial_{\alpha}, \partial_{\tau}] = 0$, it suffices to estimate
\begin{equation} \label{e2.31}
\langle \nabla_{\partial_{\alpha}}\partial_{\tau}, \partial_{\beta} 
\rangle = \langle \nabla_{\partial_{\alpha}}N, \partial_{\beta} \rangle 
+ \langle\nabla_{\partial_{\alpha}}\xi , \partial_{\beta} \rangle. 
\end{equation}
Suppose first $\alpha > 0$, $\beta > 0$, so $(\alpha ,\beta ) = (i,j)$. 
The first term in (2.31) is then $K_{ij},$ which is bounded in 
$L^{1,p}$ by (2.25), while the second term is also bounded in $L^{1,p}$ 
by (2.30), (and the $L^{2,p}$ spatial bounds on ${\bf g}_{\alpha\beta}$ 
in (2.29)). This gives uniform $L^{1,p}$ bounds on 
$\partial_{\tau}g_{ij},$ i.e. 
\begin{equation} \label{e2.32}
||\partial_{\tau}g_{ij}||_{L_{x}^{1,p}} \leq  C_{7}. 
\end{equation}
It follows of course that also 
$$||\partial_{x_{k}}\partial_{\tau}g_{ij}||_{L_{x}^{p}} \leq  C_{7}.$$ 

   The bounds on $\partial_{\tau}{\bf g}_{0\alpha}$ require more work. 
Writing $\partial_{\tau} = N+\xi $ as above, the $L^{2,p}$ spatial 
estimates above imply that $\xi ({\bf g}
_{0\alpha})$ is bounded in $L^{1,p},$ so one needs to obtain $L^{1,p}$ 
bounds on $N({\bf g}
_{0\alpha}).$ 

\medskip

 Recall that $x_{i}'  = N(x_{i}) = {\bf g}^{0i}.$ Hence $N({\bf 
g}^{0i}) = NN(x_{i}) = x_{i}''$. To obtain estimates on $x_{i}''$, 
differentiate the equation (2.17) in the $N$ direction twice. This gives
\begin{equation} \label{e2.33}
\Delta x_{i}''  = - (2\Delta' x_{i}'  + \Delta'' x_{i}). 
\end{equation}
Here, as before, all metric quantities are on the spatial slices 
$\Sigma_{\tau}.$ It has already been proved that $x_{i}'\in L^{2,p}.$ 
From the form of $\Delta' $ in (2.26), one then easily sees that 
$$\Delta' x_{i}'  \in  L^{p}. $$

 Next, one has
\begin{equation} \label{e2.34}
\Delta'' x_{i} = N\langle D^{2}x_{i}, K \rangle  - N\langle dx_{i}, 
\delta K + \tfrac{1}{2}dH \rangle. 
\end{equation}

 To estimate these terms, let $e_{\alpha}$ be a local orthonormal basis 
on $\Sigma_{\tau},$ with $\nabla_{e_{\alpha}}e_{\beta} =$ 0 at any 
fixed point in $\Sigma_{\tau}.$ Then the first term in (2.34) may be 
written 
$$N\langle D^{2}x_{i}, K \rangle  = N(D^{2}x_{i}(e_{a},e_{b}))\cdot 
K(e_{a},e_{b}) + D^{2}x_{i}(e_{a},e_{b})\cdot  N(K(e_{a},e_{b})). $$
By (2.25) and (2.7), $K_{ab}\in L^{1,p}$ and $NK_{ab}\in L^{p},$ while 
by (2.20), $D^{2}x_{i}\in L^{1,p}.$ Thus
\begin{equation} \label{e2.35}
D^{2}x_{i}(e_{a},e_{b})\cdot  N(K(e_{a},e_{b})) \in  L^{p}. 
\end{equation}
Further $N(D^{2}x_{i}(e_{a},e_{b})) = N\langle \nabla_{e_{a}}dx_{i}, 
e_{b} \rangle  = \langle \nabla_{N}\nabla_{e_{a}}dx_{i}, e_{b} 
\rangle$, so that
\begin{equation} \label{e2.36}
N(D^{2}x_{i}(e_{a},e_{b}))= \langle \nabla_{e_{a}}\nabla_{N}dx_{i}, 
e_{b} \rangle  + \langle R(N,e_{a})dx_{i}, e_{b} \rangle - \langle 
\nabla_{e_{b}}dx_{i}, \nabla_{e_{a}}N \rangle. 
\end{equation}
The curvature term in (2.36) is bounded in $L^{\infty}$, while the last 
term equals $\langle D^{2}x_{i}(e_{b}), K(e_{a}) \rangle$, which is 
bounded in $L^{\infty}$. For the first term, one has
\begin{equation} \label{e2.37}
\langle \nabla_{e_{a}}\nabla_{N}dx_{i}, e_{b} \rangle  = e_{a} \langle 
\nabla_{N}dx_{i}, e_{b} \rangle = e_{a} \langle \nabla_{e_{b}}dx_{i}, N 
\rangle = - e_{a}(K(dx_{i}, e_{b})). 
\end{equation}
Since $K\in L^{1,p},$ this term is bounded in $L^{p}.$ Combining these 
estimates, it follows that the first term in (2.34) is bounded in 
$L^{p}.$

 For the next term in (2.34), $N\langle dx_{i},dH \rangle = 
\langle \nabla_{N}dx_{i}, dH \rangle + \langle dx_{i}, \nabla_{N}dH \rangle = 
\langle \nabla_{dH}dx_{i}, N \rangle + \langle N, \nabla_{dx_{i}}dH \rangle$, 
so that
\begin{equation} \label{e2.38}
N\langle dx_{i},dH \rangle = - 2K(dH,dx_{i}),
\end{equation}
which is bounded in $L^{p}.$

 Finally, $- N\langle dx_{i}, \delta K \rangle = Ne_{a}K(dx_{i},e_{a}) - 
NK(\nabla_{e_{a}}e_{a},dx_{i})- NK(e_{a},\nabla_{e_{a}}dx_{i}).$ By 
(2.7), the latter two terms are in $L^{\infty}.$ For the first term, 
write $Ne_{a}K(dx_{i},e_{a}) = e_{a}NK(dx_{i},e_{a}) - 
(\nabla_{e_{a}}N)(K(dx_{i}, e_{b}))$. The latter term here is 
$-[K(e_{a})](K(dx_{i},e_{b}))$, which is bounded in $L^{p}$ by (2.25). 
For the first term,  since $NK(dx_{i},e_{a})$ is bounded in $L^{\infty}$, 
$e_{a}NK(dx_{i},e_{a}) = div(N(K(dx_{i})))$ is bounded in $L^{-1,p}$, since 
the derivatives $e_{a}$ are spatial. This shows that
\begin{equation} \label{e2.39}
N\langle dx_{i}, \delta K \rangle \in  L^{-1,p}. 
\end{equation}

 Thus, combining these estimates on (2.33) gives a uniform bound on 
$\Delta x_{i}''$ in $L^{-1,p}$. On the boundary $\partial\Sigma_{\tau}$, one 
has $x_{i}''  =$ 0. It then follows from elliptic regularity as before, 
(as in (2.24)), that 
$$N({\bf g}^{0i}) = x_{i}''  \in  L^{1,p}. $$
 Of course, by (2.28), $N({\bf g}^{00}) =$ 0. As above, ${\bf g}_{0i}$ 
is a rational expression in $\{g_{ij}\}$ and $\{A_{0k}\}.$ The bound 
(2.32) implies that each of these has $N$-derivative in $L^{1,p}$ and 
hence, by the same arguments as before, $N({\bf g}_{0\alpha})\in  
L^{1,p}.$ This gives uniform bounds
\begin{equation} \label{e2.40}
||\partial_{\tau}{\bf g}_{\alpha\beta}||_{L_{x}^{1,p}} \leq  C_{8}. 
\end{equation}

 This completes the estimates for the first time derivatives on spatial 
slices. In particular, all Christoffel symbols are bounded in 
$L_{x}^{1,p}.$

\medskip

{\bf $2^{\rm nd}$ Time Derivatives.}

 Finally, we obtain $L^{p}$ estimates on the $2^{\rm nd}$ time 
derivatives $\partial_{\tau}\partial_{\tau}{\bf g}
_{\alpha\beta}.$ To do this, take $\partial_{\tau}$ of the term 
$\langle \nabla_{\partial_{\alpha}}\partial_{\tau}, \partial_{\beta} 
\rangle$ in (2.31). One then obtains
$$\langle \nabla_{\partial_{\alpha}}\nabla_{\partial_{\tau}}\partial_{\tau}, 
\partial_{\beta} \rangle + \langle {\bf R}
(\partial_{\tau},\partial_{\alpha})\partial_{\tau},\partial_{\beta}\rangle.$$
The curvature term is bounded in $L^{p},$ in fact $L^{\infty}.$ (This 
uses the fact that $\xi $ is controlled, so the framing is controlled).

 Write $\nabla_{\partial_{\tau}}\partial_{\tau} = 
\Gamma_{00}^{\gamma}\partial_{\gamma}.$ Hence
$$\langle \nabla_{\partial_{\alpha}}\nabla_{\partial_{\tau}}\partial_{\tau}, 
\partial_{\beta} \rangle = \Gamma_{00}^{\gamma} \langle 
\nabla_{\partial_{\alpha}}\partial_{\gamma}, \partial_{\beta}\rangle + 
\partial_{\alpha}(\Gamma_{00}^{\gamma}) \langle \partial_{\gamma}, 
\partial_{\beta}\rangle. $$
By the first derivative estimates above, the Christoffel symbols are 
bounded in $L_{x}^{1,p}.$ The first term is a product of Christoffel 
symbols, and hence is bounded in $L_{x}^{1,p/2} \subset  
L^{np/(2n-p)},$ by Sobolev embedding. For $p > n$, (recall that $p$ is 
arbitrarily large), $L^{np/(2n-p)} \subset  L^{p}$. Thus the first term is 
bounded in $L^{p}$.

  For the second term, if $\alpha  > $ 0, then 
$\partial_{\alpha}(\Gamma_{00}^{\gamma})$ is bounded in $L^{p}$ by (2.40). 
Hence, this gives 
\begin{equation} \label{e2.41}
||\partial_{\tau}\partial_{\tau}g_{ij}||_{L^{p}} \leq  C_{9}, 
\end{equation}
uniformly in $\tau$. It remains to estimate the second time derivatives 
of ${\bf g}_{0\alpha}.$ These correspond to the second order time 
behavior of the shift vector $\xi .$

\medskip

 These estimates are the most involved, and are the only estimates 
dependent on the Einstein equations. To obtain these estimates, one 
needs to differentiate (2.17) three times in the normal $N$ direction. 
Thus, from (2.17) again,
\begin{equation} \label{e2.42}
\Delta x_{i}'''  = -3\Delta' x_{i}''  -3\Delta'' x_{i}'  -\Delta''' 
x_{i}. 
\end{equation}
Recall $x_{i}'  = {\bf g}^{0i}$ is bounded in $L^{2,p}$, while $x_{i}'' 
 = N({\bf g}^{0i})$ is bounded in $L^{1,p}.$ From previous work, it is 
then straightforward to bound the first two terms on the left in 
(2.42). 

 To see this, one has
$$\Delta' x_{i}''  = \langle D^{2}x_{i}'' , K \rangle  - \langle 
dx_{i}'' , \delta K + \tfrac{1}{2}dH \rangle, $$
which is bounded in $L_{x}^{-1,p},$ since $x_{i}''\in L_{x}^{1,p}.$ To 
estimate the term $\Delta'' x_{i}' ,$ one just replaces $x_{i}$ in the 
estimates (2.34)-(2.39) by $N(x_{i}).$ Using the fact that $N(x_{i})\in 
L^{2,p},$ one sees by checking term by term that this is bounded in 
$L^{-1,p}.$

 It remains to analyse
\begin{equation} \label{e2.43}
\Delta''' x_{i} = NN\langle D^{2}x_{i}, K \rangle  - NN\langle dx_{i}, 
\delta K + \tfrac{1}{2}dH \rangle. 
\end{equation}
To begin,
\begin{equation} \label{e2.44}
NN\langle D^{2}x_{i}, K \rangle  = \langle 
\nabla_{N}\nabla_{N}D^{2}x_{i}, K \rangle  + 2\langle 
\nabla_{N}D^{2}x_{i}, \nabla_{N}K \rangle  + \langle D^{2}x_{i}, 
\nabla_{N}\nabla_{N}K \rangle . 
\end{equation}
By the Riccati equation (2.6), $\nabla_{N}K$ is bounded in 
$L^{\infty},$ while by the estimate on (2.36), $\nabla_{N}D^{2}x_{i}$ 
is bounded in $L^{p}.$ Hence, the middle term in (2.44) is bounded in 
$L^{p}.$ For the last term, taking the $N$-derivative of the Riccati 
equation (2.6) gives 
\begin{equation} \label{e2.45}
\langle D^{2}x_{i}, \nabla_{N}\nabla_{N}K \rangle  = - 2\langle 
D^{2}x_{i}, (\nabla_{N}K)K \rangle  -  \langle D^{2}x_{i}, 
\nabla_{N}{\bf R}_{N} \rangle . 
\end{equation}
The first term in (2.45) is bounded in $L^{\infty}.$ The second 
(curvature) term in (2.45) will be analysed below.

 For the first term in (2.44), one needs to take $N$-derivatives of all 
terms following (2.35) to (2.37). The only one which is not bounded in 
$L^{-1,p}$ by previous estimates is the term
\begin{equation} \label{e2.46}
N\langle {\bf R}(N,e_{a})dx_{i},e_{b} \rangle = (\nabla_{N}{\bf 
R})(N,e_{a},dx_{i},e_{b}) + \ \ {\rm lower \ order}. 
\end{equation}
We return again to this curvature term below, and proceed with the 
second term in (2.43). 

 Again, to estimate this, one takes $N$-derivatives of the estimates in 
(2.38) to (2.39). This gives first
$$NK(dH,dx_{i}) \in  L^{p}, $$
since $NK$ is bounded. For the $\delta K$ term, modulo lower order 
terms, this is of the form
$$N\langle \nabla_{N}\nabla_{e_{a}}K(dx_{i}), e_{a} \rangle = N\langle 
\nabla_{e_{a}}\nabla_{N}K(dx_{i}), e_{a} \rangle + N\langle{\bf 
R}(N,e_{a})K(dx_{i}),e_{a} \rangle =  $$
$$= \langle \nabla_{N}\nabla_{e_{a}}\nabla_{N}K(dx_{i}), e_{a} \rangle 
+ (\nabla_{N}{\bf R})(N,e_{a},K(dx_{i}),e_{a}) = $$
$$= \langle \nabla_{e_{a}}\nabla_{N}\nabla_{N}K(dx_{i}), e_{a} \rangle 
+ (\nabla_{N}{\bf R})(N,e_{a},K(dx_{i}),e_{a}) + \langle {\bf 
R}(N,e_{a})\nabla_{N}K(dx_{i}),e_{a} \rangle; $$
the equalities here are understood to be modulo lower order terms. Modulo 
terms bounded in $L^{p},$ this may be rewritten as
\begin{equation} \label{e2.47}
\nabla_{e_{a}}(\nabla_{N}{\bf R})(N,K(dx_{i}), N, e_{a}) + 
(\nabla_{N}{\bf R})(N,e_{a},K(dx_{i}),e_{a}). 
\end{equation}

 Combining these estimates gives then uniform $L_{x}^{-1,p}$ bounds on 
all terms in (2.43), except for the four curvature terms of the 
form $(\nabla_{N}{\bf R})N.$

 We obtain bounds on the curvature terms via the contracted $2^{\rm 
nd}$ Bianchi identity on ({\bf M, g}):
$$\mbox{\boldmath $\delta$}{\bf R} = - {\bf dRic},$$
or more precisely, cf. [3, 16.3],
\begin{equation} \label{e2.48}
\mbox{\boldmath $\delta$}{\bf R}(X,Y,Z) = - {\bf dRic}(Y,Z,X). 
\end{equation}
Write, on {\bf M},
\begin{equation} \label{e2.49}
(\nabla_{N}{\bf R})N = - \mbox{\boldmath $\delta$}{\bf R} + \delta{\bf 
R} = {\bf dRic} + \delta{\bf R}, 
\end{equation}
where the divergences \mbox{\boldmath $\delta$}, $\delta $ are the 
space-time and space-like divergences on $\Sigma_{\tau}$ respectively.

 Now the space-time curvature {\bf R} is bounded in $L^{\infty},$ (in 
bounded framings). Hence the spatial divergence $\delta{\bf R}$ is 
bounded in $L_{x}^{-1,p},$ for any $p <  \infty .$ Thus, the curvature 
term (2.46) may be rewritten, modulo $L_{x}^{-1,p},$ as
$$(\nabla_{N}{\bf R})(N,e_{a},dx_{i},e_{b}) = {\bf 
dRic}(dx_{i},e_{b},e_{a}) \in  L_{x}^{-1,p}, $$
where the last estimate follows since all derivatives of {\bf Ric} are 
taken in spatial directions. Similarly, for the second curvature term 
in (2.47), one has, for the same reasons, modulo $L_{x}^{-1,p},$
$$(\nabla_{N}{\bf R})(N, e_{a},K(dx_{i}),e_{a}) = {\bf 
dRic}(K(dx_{i}),e_{b},e_{a}) \in  L_{x}^{-1,p}. $$
This leaves left the two curvature terms:
\begin{equation} \label{e2.50}
\langle D^{2}x_{i}, \nabla_{N}{\bf R}_{N} \rangle  \ {\rm and} \  
\nabla_{e_{a}}(\nabla_{N}{\bf R})(N,K(dx_{i}), N, e_{a}). 
\end{equation}
The second term is of form $\delta{\bf dRic}(N,K(dx_{i}))$, (to leading 
order), which cannot be controlled without the Einstein equation, since 
it involves differentiation in the $N$ direction. Similarly, 
\begin{equation} \label{e2.51}
\nabla_{N}{\bf R}_{N} = {\bf dRic}(N, \cdot , \cdot  ), \ \ {\rm 
modulo} \ \ L_{x}^{-1,p}, 
\end{equation}
is not controlled without the Einstein equations.

 Dropping the usual constants, the Einstein equations on ({\bf M, g}) 
are
\begin{equation} \label{e2.52}
{\bf Ric} -  \frac{{\bf R}}{2}g = {\mathcal T} , 
\end{equation}
where ${\mathcal T} $ is the stress-energy tensor. Suppose ({\bf M, g}) 
is vacuum, or more generally, suppose the stress-energy tensor 
${\mathcal T}$ satisfies
\begin{equation} \label{e2.53}
d{\mathcal T} (N, e_{a},e_{b})\in  L_{x}^{-1,p}, 
\end{equation}
where $e_{a},e_{b}$ are spatial. Since the Einstein equations and the 
bound (2.5) imply that $\nabla_{e_{a}}{\mathcal T}(N) \in  
L_{x}^{-1,p}$, (2.53) is equivalent to 
\begin{equation} \label{e2.54}
\nabla_{N}{\mathcal T}\in  L_{x}^{-1,p}, \ {\rm or} \ {\mathcal 
L}_{N}{\mathcal T} \in  L_{x}^{-1,p}, 
\end{equation}
where ${\mathcal L}_{N}$ denotes the Lie derivative in the direction 
$N$.

 Combining the bound (2.53) with the estimates obtained above on the 
terms in (2.42) then gives
\begin{equation} \label{e2.55}
\Delta x_{i}'''  = \Delta NN({\bf g}^{0i}) \in  L_{x}^{-2,p}, 
\end{equation}
since $\delta {\bf dRic}(N, K(dx_{i})) \in L_{x}^{-2,p}$, not 
$L_{x}^{-1.p}$. As before, since the coefficients of the Laplacian are 
well-controlled, (i.e. bounded in $L^{2,p}),$ and the functions 
$x_{i}''' $ have 0 boundary values, elliptic regularity gives
$$NN({\bf g}^{0i})\in L^{p}. $$
(One sees this from duality in the standard way, using the fact that 
$\Delta: L_{o}^{2,p} \rightarrow L^{p}$ is an isomorphism, cf. also 
[13]).

  Applying the linear algebra argument as before then gives the bound
\begin{equation} \label{e2.56}
\partial_{\tau}\partial_{\tau}{\bf g}_{0\alpha} \in  L^{p}. 
\end{equation}

 This completes all of the estimates on ${\bf g}_{\alpha\beta}$. We 
refer to Remark 2.1 below for further discussion on use of the Einstein 
equations. Finally, the assumptions (2.53) or (2.54) are equivalent to 
(1.13). To see this, both $T$ and $N$ lie within a compact subset of 
$T^{+}\Omega .$ Since the covariant derivative w.r.t. $T$ or $N$ 
involves only the pointwise behavior of these vector fields, one may 
replace $T$ by $N$ in (1.13), which then corresponds to (2.54).

\medskip

{\bf Step V. (Issue of Cut Points).}

  In this final step, we show that by passing to a smaller cylinder if 
necessary, of definite size within $C_{r_{o}}$, the exponential map 
$exp_{q'}$ at a suitable base point $q'$, has no future cut points. By 
the work above, this will complete the proof.

  To begin, return to the ``cone'' ${\mathcal C}$ in (2.4). Let $V$ be 
the collection of time-like unit vectors in $T_{q}^{+}{\mathcal 
D}_{r_{1}}$ for which there is a maximal geodesic $\sigma_{v}$ issueing 
from $q$ and terminating in ${\mathcal C}$. Let
\begin{equation} \label{e2.57}
\widetilde {\mathcal C} = \{sv: v \in V, s \leq r_{1}\} \subset 
T_{q}^{+}{\mathcal D}_{r_{1}},
\end{equation}
so that in particular $exp_{q}\widetilde {\mathcal C}$ contains 
${\mathcal C}$. Here, as in Step I, $r_{1}$ is chosen so that no geodesic 
$\sigma_{v}(s)$, for $v \in V$, has conjugate points with $s \leq 
r_{1}$. Of course the geodesics $\sigma_{v}(s)$ now are no longer 
necessarily maximal. Since $exp_{q}$ is of maximal rank on 
$\widetilde {\mathcal C}$, we work on the pullback 
$(\widetilde {\mathcal C}, \widetilde {\bf g})$, where
\begin{equation} \label{e2.58}
\widetilde {\bf g} = (exp_{q})^{*}{\bf g}. 
\end{equation}

  The domain $\widetilde {\mathcal C}$ is a compact connected cone in 
$T_{q}{\bf M}$, w.r.t. the vector space structure. By the Gauss Lemma, the 
straightline generators of this cone are geodesics in the 
$\widetilde {\bf g}$ metric. Let $\widetilde \tau$ denote the distance to 
the origin $\{0\}$ w.r.t. $\widetilde {\bf g}$, within 
$\widetilde {\mathcal C}$. This is now a smooth function on 
$\widetilde {\mathcal C} \setminus \{0\}$, and serves as the 
parameter for the geodesics from $\{0\}$. Inside the cutlocus of 
$exp_{q}$ on $\widetilde {\mathcal C}$, (i.e. where $exp_{q}$ is a 
diffeomorphism), $\widetilde \tau$ is just the lift of the function 
$\tau$ from (2.3), up to an additive constant. 

  The level sets $\widetilde \Sigma_{\widetilde \tau} \subset 
\widetilde {\mathcal C}$ of $\tau$ are smooth, and hence the images 
$exp_{q}\widetilde \Sigma_{\widetilde \tau}$ are smoothly immersed 
submanifolds in ${\mathcal C}$. The original Lipschitz level surface 
$\Sigma_{\tau}$ is just the part of $exp_{q}\widetilde 
\Sigma_{\widetilde \tau}$ contained in the domain $U^{\tau} = \{x: 
\tau(x) \geq \tau\} \cap {\mathcal C}$. Observe that the timelike exponential 
map $\widetilde{exp}_{0}$ of $(\widetilde {\mathcal C}, \widetilde {\bf g})$ 
based at $0$ is a diffeomorphism onto $\widetilde {\mathcal C}$; this map 
has no conjugate or cut points within $\widetilde {\mathcal C}$.  

  Thus, we are now in exactly the same situation as at the end of Step I, 
with $(\widetilde {\mathcal C}, \widetilde {\bf g})$ in place of 
$({\mathcal C}, {\bf g})$. Set $\widetilde S = (exp_{q})^{-1}({\mathcal C}) 
\subset \widetilde {\mathcal C}$, so that $\widetilde S$ is an embedded 
hypersurface in $\widetilde {\mathcal C}$. Since $exp_{q}$ is a local 
isometry of $\widetilde S$ onto $\widehat S$, one has
\begin{equation} \label{e2.59}
vol_{\widetilde g}\widetilde S \geq vol_{g}\widehat S.
\end{equation}
Hence, by the same reasoning as in Step II, the volume estimate (2.14) 
holds on the smooth hypersurfaces $\widetilde \Sigma_{\widetilde \tau}$.

\medskip

  We may thus apply the work in Steps II - IV to conclude that there is 
an $r_{o} > 0$, depending only on $C$, $D$, $c_{o}$, $v_{o}$, and an 
$r_{o}$-cylinder 
$\widetilde C_{r_{o}} \subset \widetilde {\mathcal C}$, centered on 
$\widetilde p$, on which there are coordinates 
$(\widetilde \tau, \widetilde x_{i})$ in which the metric 
$\widetilde {\bf g}$ is controlled in $L^{2,p}$ in the sense that 
(1.12) holds. 

  It remains to prove the existence of a suitable cylinder $C'$ 
``downstairs'', i.e. within $(\Omega, {\bf g})$, with these properties. 
Thus, let $\widetilde \Sigma_{\widetilde \tau}$ now denote the part of 
$\widetilde \Sigma_{\widetilde \tau}$ contained in $\widetilde 
C_{r_{o}}$. Each $\widetilde \Sigma_{\widetilde \tau}$ is an $n$-ball, 
topologically. Recall, as in Step II, that 
$\widetilde S \cap \widetilde C_{r_{o}}$ is a graph over 
$\widetilde \Sigma_{\widetilde \tau_{o}}$, for some $\widetilde \tau_{o}$. 
Hence, $\widetilde S \cap \widetilde C_{r_{o}}$ is also an 
$n$-ball topologically. By construction, both $\widetilde 
\Sigma_{\widetilde \tau_{o}}$ and $\widetilde S \cap \widetilde 
C_{r_{o}}$ have a uniform lower bound on their volume and size, i.e. 
$dist_{\widetilde g}(\widetilde p, \partial (\widetilde 
S \cap \widetilde C_{r_{o}})) \geq r_{3} > 0$, (and similarly for 
$\widetilde \Sigma_{\widetilde \tau_{o}}$). These bounds depend only on 
the initial bounds on $C$, $v_{o}$ and $D$, $c_{o}$.

  Let $U$ be the interior of the cutlocus of $exp_{q}|_{\widetilde {\mathcal C}}$. 
The domain $U$ is starshaped w.r.t. the origin $\{0\}$ in 
$T_{q}^{+}{\mathcal D}_{r_{1}}$. Further, again by construction, 
$\widetilde S \cap \widetilde C_{r_{o}}$ is contained in the closure 
$\bar U$ of $U$. If $(\widetilde S \cap \widetilde C_{r_{o}})\cap 
\partial U \neq \emptyset$, one may perturb it slightly, along the 
geodesic straight lines to $\{0\}$, to obtain an $n$-ball $\widetilde 
S' \subset \widetilde C_{r_{o}}$ with $\widetilde S' \subset U$. As 
before, the ball $\widetilde S'$ has a definite lower bound on its 
volume and its size. 

  The exponential map $exp_{q}$ now gives a diffeomorphism, in fact an 
isometry, from $(\widetilde S', \widetilde g)$ to $(S', g)$, $S' = 
exp_{q}\widetilde S'$. Let ${\mathcal D}'$ be the domain of dependence 
of $S'$ in ({\bf M, g}). This gives a globally hyperbolic region 
${\mathcal D}' \subset ({\bf M, g})$, with a simply connected Cauchy 
surface of definite size and volume. The work of Steps I - IV may now 
be applied to this situation within ({\bf M, g}) to produce a new 
cylinder $C'$, centered at $p$, satisfying the bounds (1.12).

 This completes the proof of Theorem 1.2. 
{\endproof}

 It is an open question whether Theorem 1.2 holds without an assumption 
of the form (1.13), i.e. whether there exist coordinate systems in 
which (1.12) holds under only the bounds (1.9). 

\begin{remark} \label{r2.1}
{\rm From the physical point of view, most stress-energy tensors 
${\mathcal T}$ derive from matter fields satisfying a hyperbolic system 
of PDE, of $1^{\rm st}$ or $2^{\rm nd}$ order. In such a situation, 
these equations can frequently be used to interchange a time derivative 
$\nabla_{N}{\mathcal T}$ on spatial slices, with a spatial derivative 
$\nabla_{X}{\mathcal T}$, modulo lower order terms, e.g. 
$(\nabla_{N}{\mathcal T})(X) \sim (\nabla_{X}{\mathcal T})(N)$, modulo 
lower order terms. This is exactly the process used via the $2^{\rm 
nd}$ Bianchi identity above. For instance, this is easily seen to be 
the case for electromagnetic fields, via use of the Maxwell equation 
${\bf dF} = 0$. 

  When the matter equations allow for such time-space replacement, 
modulo lower order terms, the condition (1.13) is of course not 
necessary in Theorem 1.2. }
\end{remark}

\medskip

 Next we turn to the proof of Corollary 1.3.

\medskip

 Let ({\bf M, g}) be a weakly regular space-time, satisfying the 
size conditions, and satisfying the bound (1.9) distributionally. Thus, 
the components of {\bf R}, well-defined as distributions, are in fact 
bounded in $L^{\infty}.$ 

 Any such space-time ({\bf M, g}) is a limit of a sequence of 
$(C^{\infty})$ smooth space-times $({\bf M, g}_{k})$, cf. [9, Theorem 4] 
for instance. The metrics ${\bf g}_{k}$ are obtained in the 
usual way by taking the convolution of {\bf g} with a sequence of 
smooth mollifiers. The local size conditions and local volume bound in 
(1.9), with $\delta_{o}$ in place of $\frac{1}{2}$, depend only on the 
$C^{0}$ behavior of the metric. Since the convergence to 
the limit is $C^{0},$ it follows that local size conditions and local volume 
bounds hold uniformly on the sequence $({\bf M, g}_{k})$. Similarly, the 
fact that $|{\bf R}|_{T}$ is bounded on $({\bf M, g}_{k})$ implies that the 
curvature $|{\bf R_{g}}_{k}|_{T_{k}}$ of $({\bf M, g}_{k})$ is uniformly 
bounded, for unit time-like vector fields $T_{k} \rightarrow T$, as 
$k \rightarrow  \infty .$

  It then follows from Theorem 1.2 that for any $p \in \Omega$, there are 
$r_{o}$-cylinders $((C_{r_{o}})_{k}, p_{k}) \subset\subset  
({\bf M, g}_{k})$, with $p_{k} \rightarrow p$, and coordinates 
on $(C_{r_{o}})_{k}$ in which the metric ${\bf g}_{k}$ is controlled: 
this in the sense that the bounds (1.12) hold, with the exception of 
the bound on $\partial_{0}\partial_{0}{\bf g}_{0\alpha}.$ Since the 
bounds on $C$, $v_{o}$ and $D$, $c_{o}$, hold uniformly on $({\bf M, g}_{k})$, 
and $p_{k}$ remains a bounded distance away from $\partial 
{\bf M}$, it follows that there is a limit cylinder $(C_{r_{o}}, p) 
\subset  ({\bf M, g})$ on which (1.12) holds, again except for 
the bound on $\partial_{0}\partial_{0}{\bf g}_{0\alpha}.$

\medskip

 This proves the first part of Corollary 1.3. For the second part, one 
needs to make sense of the condition (1.13) on ({\bf M, g}). It 
suffices to do this locally, i.e. on cylinders 
$C_{r_{o}} \subset  ({\bf M, g})$. Of course $\nabla_{N}{\bf Ric}$, 
or equivalently $\nabla_{N}{\mathcal T}$, is well-defined in 
$L^{-1,p}(\Omega )$, (since {\bf Ric} $\in  L^{\infty}),$ but one needs 
to define it in $L^{-1,p}(\Sigma_{\tau}).$ 

 To do this, from the above, we know that 
$g_{\tau} = {\bf g}|_{\Sigma_{\tau}}$ is in 
$L_{x}^{2,p}(\Sigma_{\tau})$, for any $\Sigma_{\tau} \subset C_{r_{o}}$. 
Let $h$ be a symmetric bilinear form in $L^{1,q}(\Sigma_{\tau}),$ of compact 
support, for a given $\tau$. Extend $h$ into $C_{r_{o}}$ by the flow of 
$N$, so that ${\mathcal L}_{N}h = 0$. Hence, $h$ is defined on all 
$\Sigma_{\tau} \subset C_{r_{o}}$. 
Formally, or alternately on the smooth approximations $({\bf M, g}_{k})$ 
of ({\bf M, g}), one has
\begin{equation} \label{e2.60}
\int_{\Sigma_{\tau}}\langle h, \nabla_{N}{\bf Ric} \rangle dV 
=\int_{\Sigma_{\tau}}N\langle h, {\bf Ric} \rangle dV 
-\int_{\Sigma_{\tau}}\langle \nabla_{N}h, {\bf Ric} \rangle dV. 
\end{equation}
Observe that $\nabla_{N}h = {\mathcal L}_{N}h + 
h(\nabla_{\partial_{a}}N,\partial_{b}) + 
h(\partial_{a},\nabla_{\partial_{b}}N) = h(K(\partial_{a}), 
\partial_{b}) + h(K(\partial_{b}), \partial_{a}).$ Since $K \in  
L^{1,p}(\Sigma_{\tau})$ and $p$ is large, it follows $\nabla_{N}h$ is 
well-defined in $L^{q}(\Sigma_{\tau})$, for any $\tau$. Hence, the second 
term in (2.60) is well-defined, (since {\bf Ric} is bounded in $L^{\infty}$).

 For the first term on the right in (2.60), one has
\begin{equation} \label{e2.61}
\int_{\Sigma_{\tau}}N\langle h, {\bf Ric} \rangle dV = 
\frac{d}{d\tau}\int_{\Sigma_{\tau}}\langle h, {\bf Ric} \rangle dV 
-\int_{\Sigma_{\tau}} \langle h, {\bf Ric} \rangle HdV. 
\end{equation}
The second term in (2.61) is well-defined, since again $H\in 
L^{1,p}\subset  C^{\alpha}.$ Thus, to define (1.13) on $C_{r_{o}}$, we 
require that the derivative
\begin{equation} \label{e2.62}
\frac{d}{d\tau}\int_{\Sigma_{\tau}}\langle h, {\bf Ric} \rangle dV, \ 
{\rm or \ equivalently}, \  \frac{d}{d\tau}\int_{\Sigma_{\tau}}\langle 
h, {\mathcal T} \rangle dV
\end{equation}
exist for all $\tau\in [-r_{o}, r_{o}]$, for any $h$ as above. Under 
this condition, the bound (1.13) on $C_{r_{o}}$ is then equivalent to a 
uniform bound on (2.62), for all $\tau\in [-r_{o}, r_{o}].$

 Given this definition of the condition (1.13) on the limit 
$C_{r_{o}},$ it follows from the same proof as in Theorem 1.2 that 
$\partial_{0}\partial_{0}{\bf g}_{0\alpha}$ is uniformly bounded in 
$L^{p}(\Sigma_{\tau})$, $|\tau| \leq r_{o}$. This completes the proof of 
Corollary 1.3.
{\endproof}

\section{Concluding Remarks.}

\setcounter{equation}{0}

 We conclude the paper with several remarks extending the validity of 
Theorem 1.2, together with some open problems.

  Theorem 1.2 gives the existence of cylinders $C$ of a definite size 
in the interior of {\bf M}, and a definite distance away from any 
boundary $\partial {\bf M}$, on which there exist coordinates in which 
the metric is controlled in $L^{2,p}.$ Of course, by rescaling up 
suitably to realize the size conditions, applying Theorem 1.2, and then 
rescaling back down, the cylinders $C$ may be chosen to be arbitrarily 
close to $\partial {\bf M}$. (This is already implicit in Corollary 1.3). 
However, the coordinates may change in the (smaller and smaller) 
cylinders as one approaches $\partial {\bf M}$.

 The main reason it is necessary to stay a definite distance away from 
$\partial {\bf M}$ in the proof is that one needs to work in globally 
hyperbolic regions, as in (2.1), which are future 1-connected. If {\bf 
M}  is globally hyperbolic and future 1-connected to begin with, then 
it is no longer necessary to stay a given distance away from $\partial 
{\bf M}$. We describe an alternate version of Theorem 1.2 in this 
context.

  Thus, suppose ({\bf M, g}) is globally hyperbolic and future 
1-connected; (for instance, ({\bf M, g}) has a simply connected Cauchy 
surface). Choose any point $q \in {\bf M}$ and time-like unit vector 
$N_{o} \in T_{q}^{+}{\bf M}$. Let $V \subset T_{q}^{+}{\bf M}$ be the 
compact cone of unit vectors forming a fixed angle $\theta$ with 
$N_{o}$, and let ${\mathcal C} = exp_{q}\tau v$, $v \in V$, be the 
corresponding ``cone'' in ({\bf M, g}). Also, let $N$ be the 
corresponding vector field tangent to the geodesics $\sigma_{v}(\tau) = 
exp_{q}(\tau v)$, and let $\Sigma_{\tau}$ be the level sets of $\tau$ in 
${\mathcal C}$. We make the following assumptions: there exist 
constants $C < \infty$ and $v_{o} > 0$ such that
\begin{equation} \label{e3.1}
|{\bf R}|_{N} \leq C, \ {\rm within} \ {\mathcal C},
\end{equation}
\begin{equation} \label{e3.2}
vol\Sigma_{\tau} \geq v_{o}\cdot (diam\Sigma_{\tau})^{n}.
\end{equation}
Although (3.2) is understood to hold for all $\tau \leq r_{1}$, for a 
fixed ${r_{1}}$, the methods used in Step II show that it suffices to 
assume (3.2) holds for a fixed $\tau_{o} > 0$, where the size 
$\tau_{o}$ depends only on $C$ in (3.1). The bound (3.2) is of course 
scale-invariant.

  We then have the following result, valid at least up to a point in 
$\partial {\bf M}$.

\begin{corollary} \label{c3.1}
Let ({\bf M, g}) be a globally hyperbolic, future 1-connected 
space-time satisfying the assumptions (3.1)-(3.2). Suppose also ({\bf 
M, g}) is vacuum, or more generally, satisfies (1.13). Then there exist 
small constants $d_{o} > 0$, and $r_{o} > 0$, depending only on $C$, 
$v_{o}$, $N_{o}$ and $\theta$, such that if
$$\sigma_{N_{o}}(\tau) \in {\bf M}, \ \forall \tau < d_{o},$$
then there exist coordinates $(\tau, x_{i})$ on the cylinder
\begin{equation} \label{e3.3}
C_{r_{o}}(d_{1}) = D_{p}(r_{o}d_{o}) \times [\tfrac{d_{1}}{2}, d_{1}] 
\subset {\bf M},
\end{equation}
in which the metric satisfies the bounds (1.12). Here $d_{1}$ is the 
largest value such that $d_{1} \leq d_{o}$ and $C_{r_{o}}(d_{1})$ is 
contained in {\bf M}.
\end{corollary}

  The proof of this result is exactly the same as that of Theorem 1.2. 
In fact it is simpler, since the issues of global hyperbolicity and 
future 1-connected are assumed, and one works with $N$ in place of the 
vector field $T$ from Definition 1.1. Thus, for $d_{o}$ small as above, 
and for $C(V)$ the cone on $V$ in $T_{q}{\bf M}$, the exponential map 
$exp_{q}$ restricted to $(C(V) \setminus \{0\}) \cap B_{0}(d_{1})$ is a 
diffeomorphism onto $({\mathcal C} \cap B_{q}(d_{1})) \setminus {q}$, 
where $B_{q}(d_{1}) = exp_{q}(B_{0}(d_{1}))$. The domain ${\mathcal C} 
\cap B_{q}(d_{1})$ plays exactly the same role as ${\mathcal C}$ in 
Step I. All the estimates of Steps II-IV then proceed just as before and 
are uniform for $\tau \in [\frac{d_{1}}{2}, d_{1}]$. 
{\endproof}

\begin{remark} \label{r 3.2.}
  {\rm One may also derive a version of Theorem 1.2, (or Corollary 
3.1), without the lower volume bound in (1.9). Thus, one has a uniform 
spatial curvature bound (2.9), either on the slices 
$\Sigma_{\tau} \subset {\mathcal C}$ downstairs, or on the slices 
$\widetilde \Sigma_{\widetilde \tau} \subset \widetilde {\mathcal C}$ 
upstairs in $T_{q}{\bf M}$. (The bound (2.9) does not require any 
apriori volume bound). If the volume $\Sigma_{\tau}$ or 
$\widetilde \Sigma_{\widetilde \tau}$ is very small, then the injectivity 
radius of $g_{\tau}$ is very small, i.e. the spatial metric $g_{\tau}$ is 
highly collapsed in the sense of Cheeger-Gromov. However, (as before), 
the Rauch comparison theorem implies that the intrinsic exponential map 
on $\Sigma_{\tau}$ or $\widetilde \Sigma_{\widetilde \tau}$ is still of 
maximal rank on geodesic balls of a definite size, (depending only on the 
curvature bound). Suppose first one works in the situation of 
$\Sigma_{\tau} \subset {\mathcal C}$. Then just as before in time-like 
directions, one can lift the metric $g_{\tau}$ up to the tangent space by 
pulling back by its exponential map, i.e. consider the metric 
$\widetilde g_{\tau} = (exp_{p_{\tau}})^{*}g_{\tau}$ defined on balls in 
$T_{p_{\tau}}\Sigma_{\tau}$. The metric $\widetilde g_{\tau}$ has a 
uniform lower bound on the volumes of small balls. All the arguments in 
Steps II-IV can then be carried out as before on this ``unwrapped'' 
space-time and the corresponding unwrapped cylinder $\widetilde 
C_{r_{o}}.$ This gives coordinates $(\widetilde \tau , \widetilde 
x_{i})$ for $\widetilde C_{r_{o}}$ on which the metric $\widetilde {\bf 
g}$ satisfies the bounds (1.12). The same procedure holds when working 
with $\widetilde \Sigma_{\widetilde \tau}$. These coordinates upstairs 
within $T_{q}{\bf M}$ give then ``multi-valued'' coordinates downstairs in 
({\bf M, g}). }
 
\end{remark}

 To conclude, we mention two open problems. First, for certain 
purposes, the $L^{\infty}$ bound on the curvature in (1.11) may be 
viewed as too strong. It would be of interest to know if a version of 
Theorem 1.2 holds with suitable $L^{p}$ bounds on $|{\bf R}|_{T}$, $p > 
n/2$, in place of $L^{\infty}$ bounds.

 Second, it would be very interesting if the Einstein equations could 
be used to remove the dependence of these results on bounds on the full 
curvature, i.e. if bounds on the full curvature could be replaced by 
bounds on the Ricci curvature, possibly introducing other hypotheses 
not related to curvature. This seems to be a challenging problem; cf. 
[1] for some further discussion.

\bibliographystyle{plain}

\begin{thebibliography}{WWW}
\footnotesize


\bibitem [1]{1} Anderson, M.: Cheeger-Gromov theory and applications to 
general relativity, Proc. 2002 Carg\`ese School on General Relativity, 
(to appear), gr-qc/0208079. 

\bibitem [2]{2} Beem, J, Ehrlich, P and Easley K.: {\it Global 
Lorentzian Geometry}, $2^{\rm nd}$ Edition, Marcel Dekker, New York, 
(1996).

\bibitem [3]{3} Besse, A.: {\it Einstein Manifolds}, Ergebnisse der 
Mathematik Series, {\bf 3:10}, Springer Verlag, Berlin, (1987). 

\bibitem [4]{4} Clarke, C.J.S.: Singularities in globally hyperbolic 
space-time, Comm. Math. Phys., {\bf 41}, (1975), 65-78.

\bibitem [5]{5} Clarke, C.J.S.: Local extensions in singular 
space-times II, Comm. Math. Phys., {\bf 84}, (1982), 329-331. 

\bibitem [6]{6} Clarke, C.J.S.: {\it The Analysis of Space-Time 
Singularities}, Cambridge Lecture Notes in Physics, 1, Cambridge Univ. 
Press, London, (1993).

\bibitem [7]{7} Ellis, G.F.R. and Schmidt, B: Singular space-times, 
Gen. Rel. and Gravitation, {\bf 8}, (1977), 915-953.

\bibitem [8]{8}  Gantmacher, F.: {\it The Theory of Matrices}, vol. 1, 
Chelsea Publishing Co, New York, (1960).

\bibitem [9]{9} Geroch, R. and Traschen, J.:, Strings and other 
distributional sources in general relativity, Phys. Review {\bf D36}, 
(1987), 1017-1031.

\bibitem [10]{10} Gilbarg, D. and Trudinger, N.: {\it Elliptic Partial 
Differential Equations of Second Order}, $2^{\rm nd}$ Edition, Springer 
Verlag, Berlin, (1983).

\bibitem [11]{11} Hawking, S.W. and Ellis, G.F.R.: {\it The Large Scale 
Structure of Space-Time}, Cambridge Univ. Press, London, (1973).

\bibitem [12]{12} Jost, J. and Karcher, H.: Geometrische Methoden zur 
Gewinnung von a-priori Schranken f\"ur harmonische Abbildungen, 
Manuscripta Math., {\bf 40}, (1982), 21-71.

\bibitem [13]{13} Lions, J.L. and Magenes, E.: {\it Non-Homogeneous 
Boundary Value Problems and Applications}, I, Grundlehren Series, {\bf 
130}, Springer Verlag, Berlin, (1966).

\bibitem [14]{14} Tipler, F., Clarke, C.J.S. and Ellis, G.F.R: {\it 
Singularities and horizons: a review article}, in General Relativity 
and Gravitation, vol. 2, A. Held, (Ed.), Plenum Press, New York, 
(1980), 87-206.

\end{thebibliography}

\medskip
\begin{center}
September, 2002
\end{center}

\noindent
\address{Department of Mathematics\\
S.U.N.Y. at Stony Brook\\
Stony Brook, N.Y. 11794-3651\\}
\noindent
\email{anderson@math.sunsyb.edu}

\end{document}